# Securing Medical Images by Watermarking Using DWT-DCT-SVD

Nilesh Rathi[#1], Ganga Holi[2]

[#]M Tech, IV semester[#1], Associate Professor[#2],
ISE Department, P E S Institute of Technology,
Bangalore, INDIA.
[1]nileshrathi345@gmail.com  [2]ganga_h@pes.edu

*Abstract*— Telemedicine is well known application where enormous amount of medical data need to be transferred securely over network and manipulate effectively. Security of digital data, especially medical images, becomes important for many reasons such as confidentiality, authentication and integrity. Digital watermarking has emerged as a advanced technology to enhance the security of digital images. The insertion of watermark in medical images can authenticate it and guarantee its integrity. The watermark must be generally hidden does not affect the quality of the medical image. In this paper, we propose blind watermarking based on Discrete Wavelet Transform (DWT), Discrete Cosine Transform (DCT) and Singular Value Decomposition (SVD), we compare the performance of this technique with watermarking based DWT and SVD. The proposed method DWT, DCT and SVD comparatively better than DWT and SVD method.

*Keywords*— Digital Watermarking, Medical images, Discrete Wavelet Transform (DWT) Discrete Cosine Transform (DCT) Singular Value Decomposition (SVD), Security.

## I. INTRODUCTION

The rapid development of internet technology and multimedia in every field leads to availability of digital data to the public. Internet has been spread in many applications like telemedicine, online-banking, teleshopping etc.

In telemedicine application special safety, confidentiality and integrity is required for medical images, because critical judgment is done on medical images, which leads to the proper treatment. The main aspects of security concerned with medical data are confidentiality, authentication, integrity and availability images, because critical judgment is done on medical images, which leads to the proper treatment.

Cryptography provides several solutions but alone have become insufficient to cover all safety aspects of the processing and transfer of images and digitized medical record. Digital watermarking is emerging technique for copyright protection and authentication for these medical images, which includes the embedding and extraction process.

In embedding process some secret information is embedded into medical images. Extraction process deals with the extraction of secret message, which is embedded in the medical image. The secrete image will be used for authenticating the cover image for the security purpose. The image tampering can be detected by extracting the secrete image from the watermarked image. If the extracted secrete image is not matching with the database image, then receiver will come to know that there has been some kind of tampering. In case of medical image, the receiver will analyze the medical data after authentication. If the medical image is tampered, then receiver will discard these medical image . Typical diagram for medical image watermarking is given in figure 1.

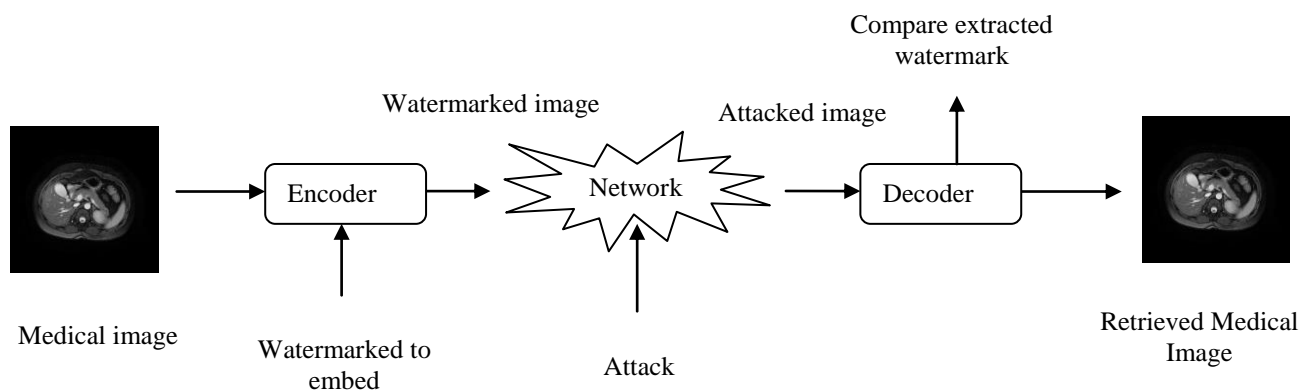

Figure 1: Block Diagram of Medical Image Watermarking

According to watermark embedding process, watermarking techniques are classified into two different domains.

- Spatial Domain: The spatial-domain watermark insertion manipulates image pixels. However, the spatial-domain watermark insertion is simple and easy to implement, it is weak against various attacks and noise.
- Transform domain: The transform-domain watermark insertion is based on the transform coefficients of cover image. It is more robust against attacks. Discrete Cosine Transform (DCT), Discrete Wavelet Transform (DWT), and Discrete Fourier Transform (DFT) are three popular methods in transform domain.

Properties of watermarking:

An effective digital watermarking algorithm must have number of properties.

- Imperceptibility: The basic requirement of digital watermarking is to have the watermarked image should look alike as the original image. This confirms there is not much degradation on the original image.
- Robustness: The watermark must resist changes due to some unintentional attacks or legitimate and illegitimate attacks.
- Capacity: Capacity of the watermarking system describes embedding of maximum amount of watermark information
- Invertibility: The digital watermarking system describes the possibility of generating original data during the extraction process of watermark.

In this paper, we propose watermarking scheme based on DWT-SVD and DWT-DCT-SVD, evaluate the results of both. The paper is organized as follows: section 2 presents the related work on watermarking. Section 3 presents methodology used for watermarking. In section 4, we propose a DWT–SVD and DWT-DCT-SVD based watermarking algorithms. Section 5 provides the details of experiments performed with their result, finally conclusions are drawn.

## II. RELATED WORK

Numbers of medical image watermarking schemes reported in this related work, to address the issues of medical information security, and authentication. Wakatani[1] proposed a medical image watermarking, in order not to compromise with the diagnosis value, it avoids embedding watermark in the ROI. In this algorithm watermark to be embed is firstly compressed by progressive coding algorithm such as Embedded Zero Tree Wavelet (EZW). Embedding process is done by applying Discrete Wavelet Transform (DWT). Extraction of watermark is reverse of embedding process. The major drawback of this algorithm is ease of introducing copy attack on the non-watermarked area. Yusuk Lim et al.[2] reported a web-based image authentication system for CT scan images. This technique considers the principal of verifying the integrity and authenticity of medical images. In this approach, the watermark is preprocessed by using 7 most significant bit-planes except least significant bit (LSB) plane of cover medical image, as a input to the hash function. This hash function generates binary value of 0 or 1 using secrete key, which is then embedded in LSB bit of cover image to get watermarked image. Hemin Golpira[3] et al. reported reversible blind watermarking. In which embedding process, firstly by applying Integer Wavelet Transform (IDWT) image is decomposed into four sub bands. By selecting two points, called thresholds, according to the capacity required for the watermark data, watermark is embedded. To get watermarked image Inverse Integer Wavelet Transform (IIDWT) is applied. In the extraction process, all of these stages are performed in reverse order to extract watermark as well as host image.

Ghazy *et al* (2007)[9] divided the image into non-overlapping blocks and then applied SVD to these blocks. Singular values of these blocks were used to embed the watermark. This scheme gave good results against compression, filtering, noise addition but failed against cropping and geometric attacks. Bhandari *et al* (2005)[11] used spread spectrum (SS) along with SVD to increase the robustness of watermarking scheme. Used two watermarks to embed, one was embedded using SS and other by pure SVD. This complementary technique covers wide range of attacks and also non-blind in nature.

Quan & Qingsong[15] proposed hybrid method based on DCT and SVD. In this DCT is applied to cover image and separated into frequency bands. SVD of DCT transformed are needed to modify with singular values of watermark to generate watermarked image. Ganic & Ahmet Eskicioglu[14] presented watermarking scheme using DWT frequency domain. The following section gives brief description about the techniques frequently used in the watermarking techniques. The proposed method also uses the same technique for securing medical images.

## III. METHODOLOGY

*A. Singular Value Decomposition (SVD):*

SVD is based on a theorem from linear algebra which says that a rectangular matrix A can be broken down into the product of three matrices - an orthogonal matrix U, a diagonal matrix S, and the transpose of an orthogonal matrix V. The theorem is usually presented something like this:

$$A = U*S*V^T \quad (1)$$

where U and V are the unitary matrices such that $U*U^T = I$ and $V*V^T = I$, where, I is an Identity matrix, S is a diagonal matrix.

Diagonal elements of S are the singular values and they satisfy the following property

$$s(1,1) > s(2,2) > s(3,3) > \ldots\ldots\ldots > s(n,n) \quad (2)$$

SVD is popular for the watermarking because:
- Few singular values can represent large portion of signal energy,
- SVD can be applied to square and rectangular images,
- The SV's (singular values) of an image have very good noise immunity, i.e., SV's do not change significantly when a small perturbation is added to an image intensity values,
- SV's represent intrinsic algebraic properties.

### B. Discrete Cosine Transform (DCT):

DCT expresses a finite sequence of data points in terms of a sum of cosine functions oscillating at different frequencies. Important to many applications like lossy compression of audio and images also to spectral methods for the numerical solution of partial differential equations.

For DCT with block size (M x N), the connection between the spatial domain image pixels X(i, j) and the transform domain coefficients Y (u, v) is as follows

$$Y(u,v) = \frac{2c(u)c(v)}{\sqrt{MN}} \sum_{i=0}^{M-1} \sum_{j=0}^{N-1} X(i,j) \cos\left[\frac{(2i+1)u\pi}{2M}\right] \cos\left[\frac{(2j+1)v\pi}{2N}\right] \quad (3)$$

Where u = 0, 1,….,M – 1, v = 0, 1, ….,N - 1, and

$$c(k) \begin{cases} \frac{1}{\sqrt{2}}, & if\ k = 0; \\ 1, & otherwise \end{cases}$$

### C. Discrete Wavelet Transform (DWT):

DWT is a multi-resolution decomposition of a signal. It is a new signal analysis theory and is a "time-frequency" method i.e. it captures both frequency and location information (location in time). The basic idea in the DWT for a one dimensional signal is the following. A signal is split into two parts, usually high frequencies and low frequencies. The edge components of the signal are largely contained to the high frequency part. The low frequency part is split again into two parts of high and low frequencies. This process is continued an arbitrary number of times, which is usually determined by the application at hand. Furthermore, from these DWT coefficients, the original signal can be reconstructed. This reconstruction process is called the inverse DWT (IDWT). The DWT and IDWT can be mathematically stated as follows

Let following be a low-pass and a high-pass filter, respectively.

$$H(w) = \sum_{k} h_k e^{-jkw} \text{ and } G(w) = \sum_{k} g_k e^{-jkw} \quad (4)$$

The above equation satisfies a certain condition for reconstruction to be stated later. A signal, x[n] can be decomposed recursively as

$$c_{j-1,k} = \sum_{n} h_{n-2k} c_{j,n}$$

$$d_{j-1,k} = \sum_{n} g_{n-2k} c_{j,n} \quad (5)$$

Furthermore, the signal x[n] can be reconstructed from its DWT coefficients recursively

$$c_{j,n} = \sum_{k} h_{n-2k} c_{j-1,k} + \sum_{k} g_{n-2k} d_{j-1,k} \quad (6)$$

For 2-D images, applying DWT corresponds to processing the image by 2-D filters in each dimension. The filters divide the input image into four non-overlapping multi-resolution sub bands LL1, LH1, HL1 and HH1. The sub-band LL1 represents the coarse-scale DWT coefficients while the sub-bands LH1, HL1 and HH1 represent the fine-scale of DWT coefficients. To obtain the next coarser scale of wavelet coefficients, the sub-band LL1 is further processed until some final scale N is reached.

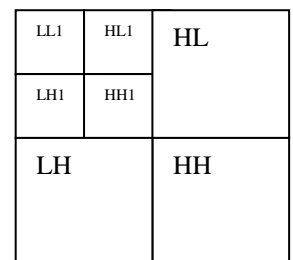

Figure 2: Sub-band decomposition using a 2D-DWT with 2 transformation levels.

### IV. PROPOSED SYSTEM

#### A. DWT-SVD:

We initially implement a simple watermarking scheme which is based on cascading DWT with SVD for comparison with the DWT, DCT and SVD method. DWT decomposes the image into four frequency bands: LL, HL, LH, and HH band. LL band represents low frequency, HL and LH represent middle frequency and HH represents high frequency band respectively.

LL band represents approximate details, HL band gives horizontal details, LH provides vertical details and HH band highlights diagonal details of the image. In this proposal, we select HH band to embed the watermark because it contains the finer details and contributes insignificantly to the image energy. Hence watermark embedding will not affect the perceptual fidelity of cover image. Moreover, high energy LL band coefficient cannot be tweaked beyond certain point as it will severely impact perceptual quality.

The proposed scheme is replaces singular values of the HH band with the singular values of the watermark. Replacement of the singular values will not affect perceptual quality of image.

*1) Watermark Embedding Algorithm:*
 i. Watermark W is decomposed using SVD
 $$W = U_w * S_w * V_w^T. \quad (7)$$
 ii. Apply DWT and decompose medical image into four sub-bands: LL, HL, LH, and HH.
 iii. Apply SVD to HH band.
 $$H = U_H * S_H * V_H^T. \quad (8)$$
 iv. Replace the singular values of the HH band with the singular values of the watermark.
 v. Apply inverse SVD to obtain the modified HH band.
 $$H' = U_H * S_w * V_H^T. \quad (9)$$
 vi. Apply inverse DWT to produce the watermarked medical image.
 vii. Pass HH band to extract original medical image.

*2) Watermark Extracting Algorithm:*
 i. Using DWT, decompose the watermarked image into four sub-bands: LL, HL, LH, and HH.
 ii. Apply SVD to HH band.
 $$H = U_H * S_H * V_H^T \quad (10)$$
 iii. Extract the singular values from HH band.
 iv. Construct the watermark using singular values and orthogonal matrices Uw and Vw obtained using SVD of original watermark.
 $$W_E = U_w * S_H * V_w^T \quad (11)$$
 v. Reconstruct retrieved medical image by inverse DWT using HH band and rest band of watermarked image.

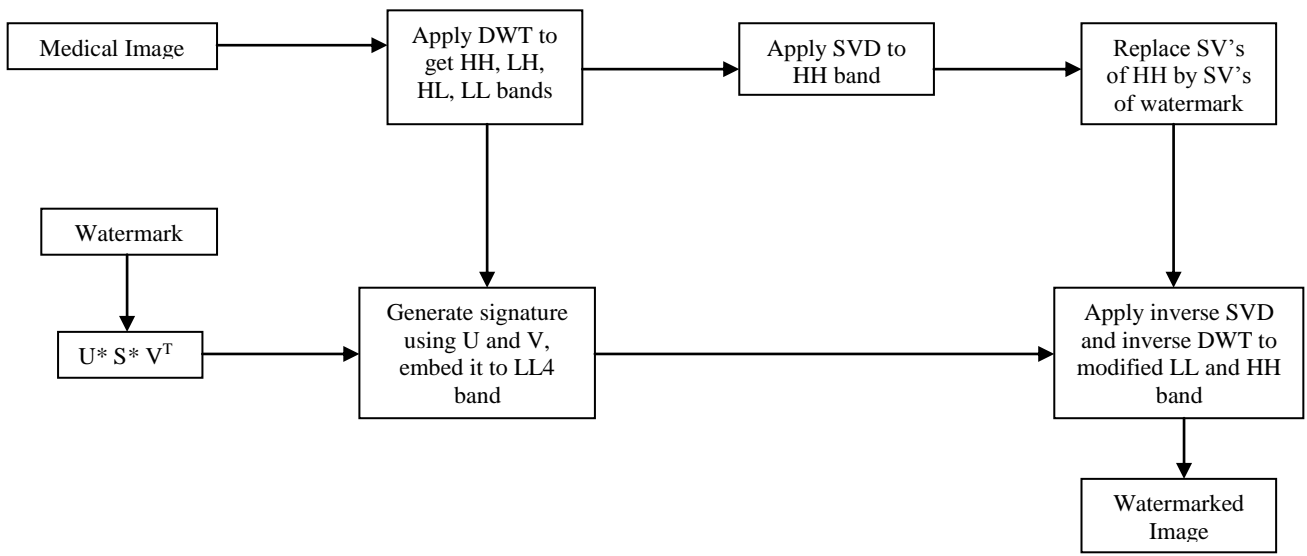

Figure 3: Block diagram of proposed scheme at encoder side for DWT-SVD

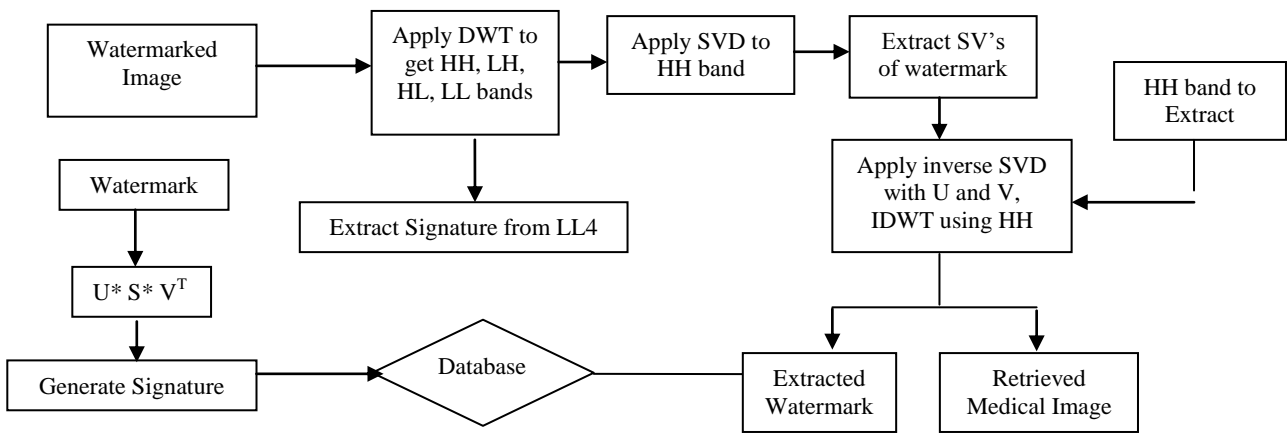

Figure 4: Block diagram of proposed scheme at decoder side for DWT-SVD

The block diagrams for the encoding and decoding algorithms are shown in Figures (3) & (4).

B.  *DWT-DCT-SVD:*

Here we propose watermarking scheme based all cascading all three DWT, DCT and SVD technology. DWT decomposes the image into four frequency bands: LL, HL, LH, and HH band. Take HH band of these four bands and apply DCT to chosen sun band. Consider it as B matrix. The proposed scheme replaces singular values of the B matrix with the singular values of the watermark.

While extracting the watermark after applying DWT to watermarked image again DCT applied to HH band of watermarked image to get cosine matrix B'. Further from B' SV's can be extracted of the watermarked image.

And retrieved medical image can be obtained by applying inverse DWT using HH band from sender and rest of bands of watermarked image. The block diagram for the Embedding and extracting watermark in medical image using DWT, DCT and SVD are given in the Figures(5) & (6) respectively.

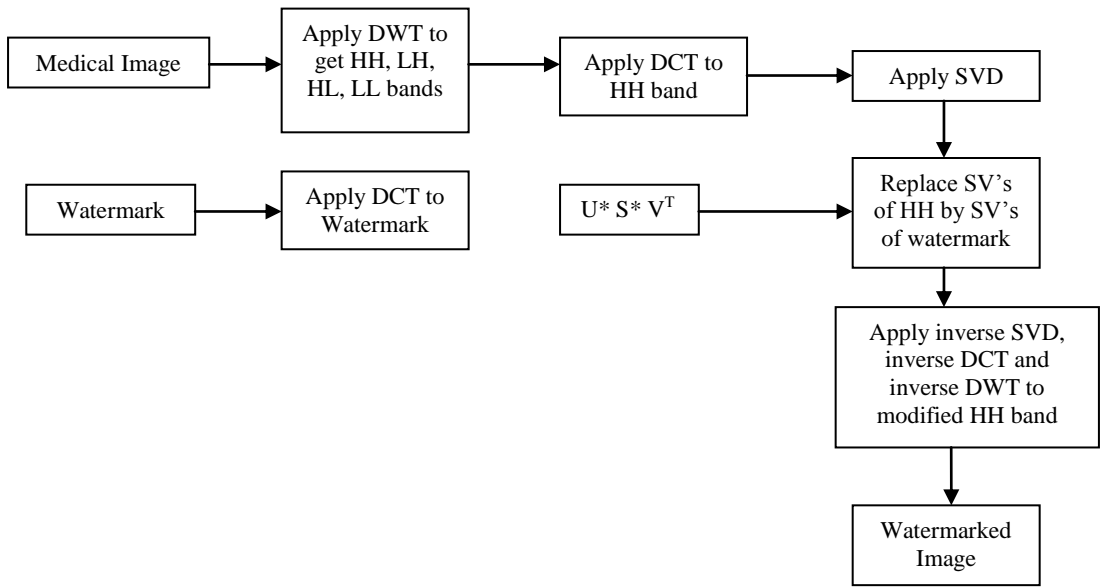

Figure 5: Block diagram of proposed scheme at encoder side for DWT-DCT_SVD

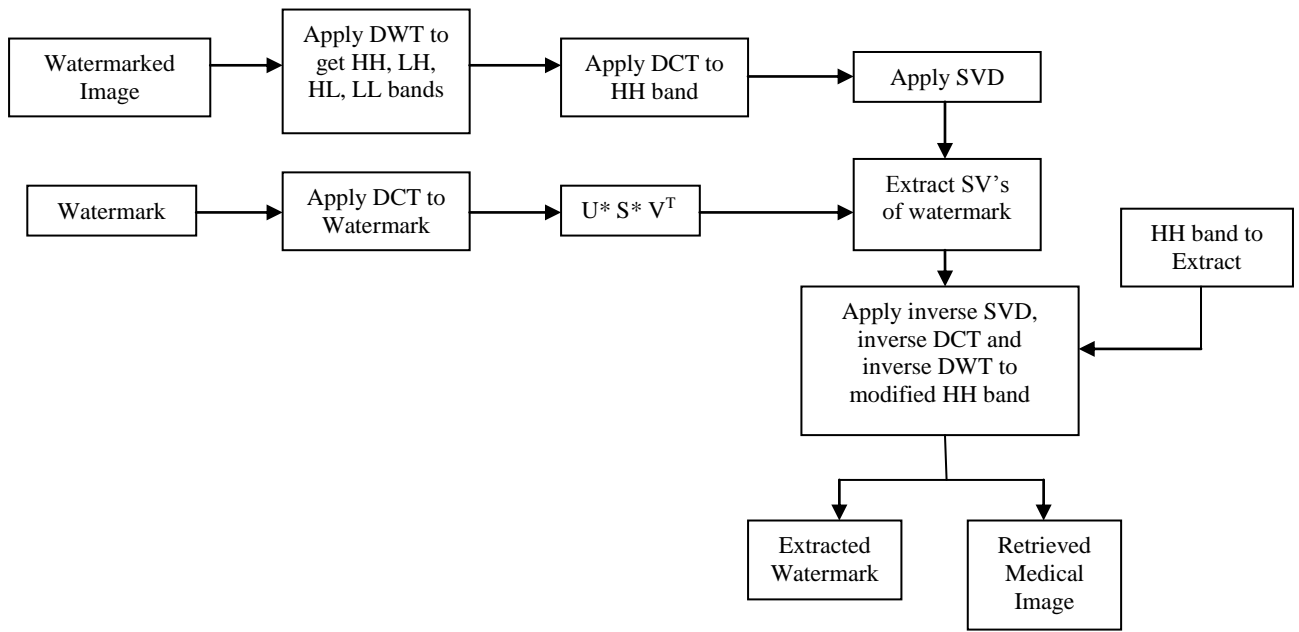

Figure 6: Block diagram of proposed scheme at decoder side for DWT-DCT_SVD

*1) Watermark Embedding Algorithm:*
  i. Watermark W is decomposed using SVD
     W = Uw *Sw *Vw^T. (12)
  ii. Apply DWT and decompose medical image into four sub-bands: LL, HL, LH, and HH.
  iii. Apply DCT to HH band and get decompose same using SVD technique
     H = U_H *S_H *V_H^T. (13)
  iv. Replace the singular values of above band with the singular values of the watermark.
  v. Apply inverse SVD and inverse DCT to obtain the modified HH band.
     H' = U_H *Sw *V_H^T. (14)
  vi. Apply inverse DWT to produce the watermarked medical image.
  vii. Pass HH band to extract original medical image.

*3) Watermark Extracting Algorithm:*
  i. Using DWT, decompose the watermarked image into four sub-bands: LL, HL, LH, and HH.
  ii. Apply DCT to HH band and further SVD to same.
     H = U_H *S_H *V_H^T (15)
  iii. Extract the singular values from HH band.
  iv. Construct the watermark using singular values and orthogonal matrices Uw and Vw obtained using SVD of original watermark.
     W_E = Uw *S_H *Vw^T (16)
  v. Reconstruct retrieved medical image by inverse DCT, inverse DWT using HH band and rest band of watermarked image.

V. RESULTS AND DISCUSSION

For above proposed method have been experimented on dataset of around 1000 medical images with different types of logos. Images and logos are gray scale are considered for experimentation of size 512 x 512. Above method can be applicable to any kind of medical image such as CT scan, MRI, X-ray etc. The experimental results are shown in Figure (7)-(14) without attack and with attack.

To ensure the reliability and quality of the watermarked image, the performance of watermarking is calculated, which measured in terms of perceptibility. There are two method of calculating the performance measure.

- Mean Square Error (MSE): It is simplest function to measure the perceptual distance between watermarked and original image. MSE can be defined as:

$$MSE = \frac{1}{n}\sum_{i}^{n}(I' - I)^2 \quad (17)$$

- Peak Signal to Noise Ratio (PSNR): It is used to measure the similarity between images before and after watermarking.

$$PSNR = 10 \log_{10} \frac{maxI}{MSE} \quad (18)$$

Normal Correlation values (NC) compares the input and output image and provides result. If NC value is '1' concludes that no attack occurred on image, else if NC value is less than '1' concludes an external attack occurred on image. Value ranges between '0' to '1' gives tampering or distorted factor of image. For the less value of NC declares more amount of distortion on image.

In our experiments and results we have compared results of watermarking techniques as DWT-DCT-SVD and DWT-SVD. After comparing NC values of both methods it has been found that DWT-DCT-SVD gives more robustness to image as compared to DWT-SVD.

A. *Watermarking using DWT-DCT-SVD:*

*1) Without Attack:*

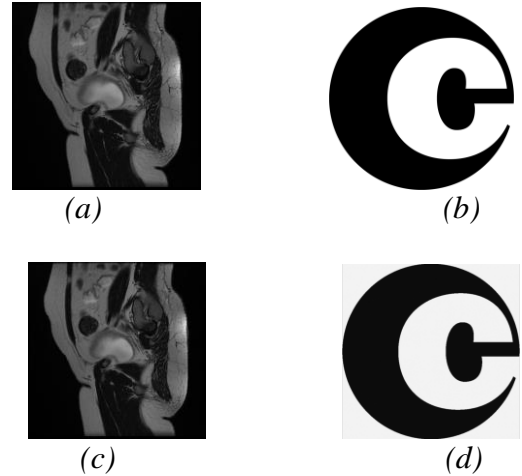

*(a)* *(b)*

*(c)* *(d)*

Figure 7: (a) Medical Image (b) Watermark (c)Watermarked Image (d) Extracted Watermark

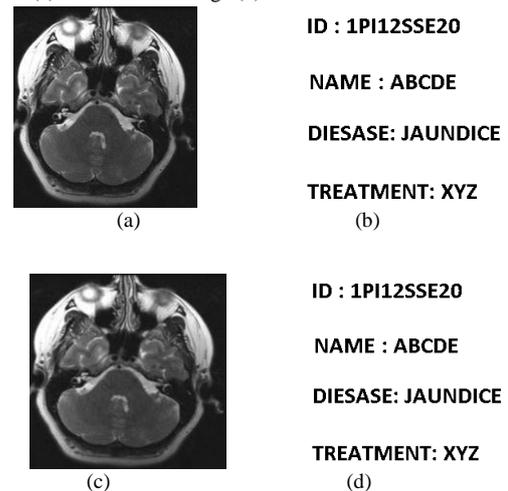

(a) (b)

(c) (d)

Figure 8: (a) Medical Image (b) Watermark (c) Watermarked Image (d) Extracted Watermark

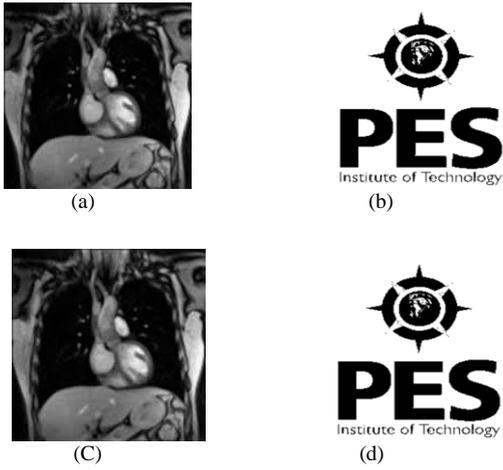

(a)      (b)

(C)      (d)

Figure 9: (a) Medical Image (b) Watermark
(c) Watermarked Image (d) Extracted Watermark

Table I: PSNR and CN for without any attack

| SEQUENCE | PSNR | NC |
|---|---|---|
| 1 | 47.4032 | 1 |
| 2 | 47.9190 | 1 |
| 3 | 47.5898 | 1 |
| 4 | 49.2799 | 1 |
| 5 | 50.8055 | 1 |
| 6 | 46.6372 | 1 |
| 7 | 47.0819 | 1 |
| 8 | 46.7991 | 1 |
| 9 | 48.1555 | 1 |
| 10 | 49.2543 | 1 |

*2) With Attack:*
*i. Median Attack:*

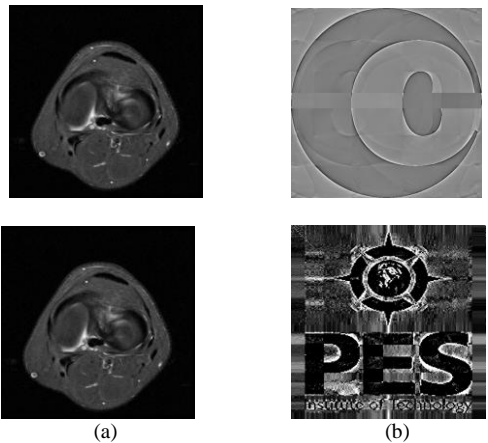

(a)      (b)

Figure 10: (a) Extracted medical images (b) Extracted Watermarks

*ii. Noise Attack:*

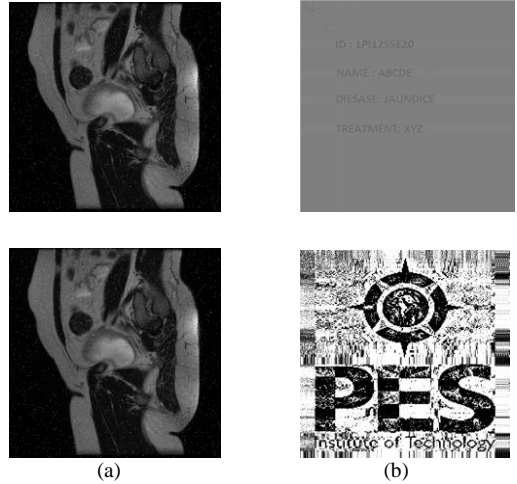

(a)      (b)

Figure 11: (a) Extracted medical images (b) Extracted Watermarks

*iii. Rotation Attack:*

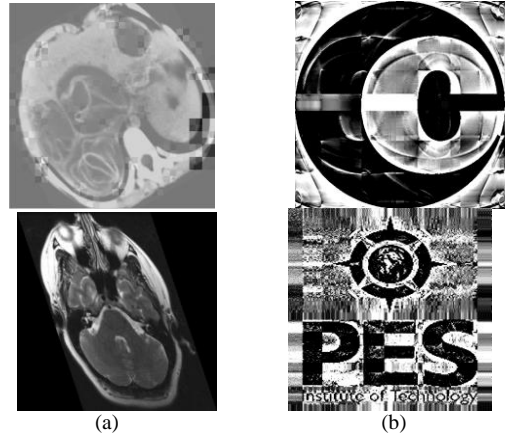

(a)      (b)

Figure 12: (a) Extracted medical images (b) Extracted Watermarks

*iv. Shear Attack:*

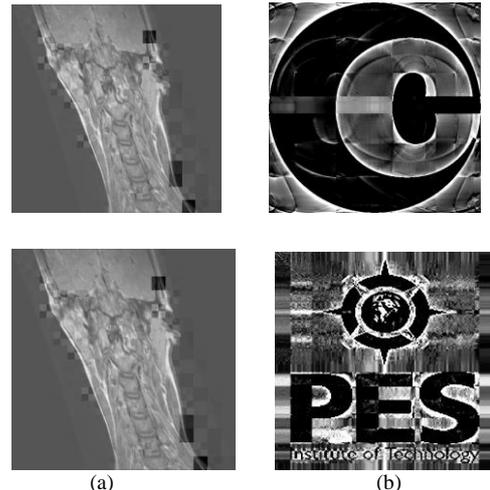

(a)      (b)

Figure 13: (a) Extracted medical images (b) Extracted Watermarks

*v.   Crop Attack:*

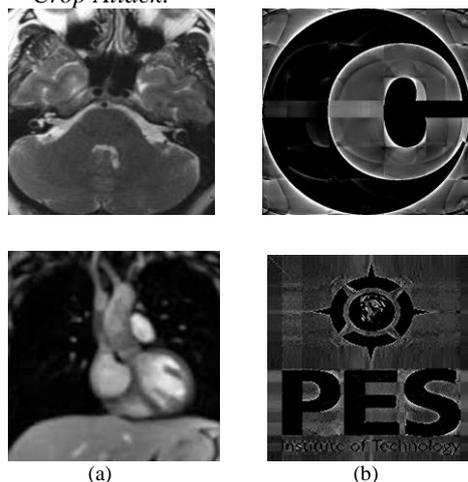

(a)              (b)
Figure 14: (a) Extracted medical images (b) Extracted Watermarks

Table II: PSNR values and NC for different types of attack

| PSNR VALUES | ATTACK | NC |
|---|---|---|
| 50.1573 | MEDIAN | 0.5712 |
| 51.5093 | NOISE | 0.4927 |
| 50.4167 | ROTATION | 0.5818 |
| 53.2607 | SHEAR | 0.6002 |
| 52.2350 | CROP | 0.7110 |
| 46.6372 | MEDIAN | 0.6162 |
| 47.0819 | NOISE | 0.5192 |
| 47.7891 | ROTATION | 0.5825 |
| 48.1555 | SHEAR | 0.6582 |
| 49.2543 | CROP | 0.8012 |

*B.   Watermarking using DWT-SVD:*

.   Table III: PSNR and NC for DWT-SVD without attack

| SEQUENCE | PSNR | NC |
|---|---|---|
| 1 | 44.2012 | 1 |
| 2 | 45.1190 | 1 |
| 3 | 42.5848 | 1 |
| 4 | 51.2546 | 1 |
| 5 | 50.8055 | 1 |
| 6 | 46.6372 | 1 |
| 7 | 43.0819 | 1 |
| 8 | 46.7991 | 1 |
| 9 | 48.1555 | 1 |
| 10 | 53.2543 | 1 |

Figure (7).(a), (8).(a) & (9).(a) show original medical Images, Figure (7).(b), (8).(b) & (9).(b) show Watermark images used to embed for security. Figure (7).(c), (8).(c) & (9).(c) show watermarked images and Figure (7).(d), (8).(d) & (9).(d) are the extracted medical images.

Figures (10) to Figures (14) are results of the proposed method which are showing modified watermarked image after extraction. These are generated by applying different attacks. Proposed method identifies the attacks effectively and alters the watermark image. The Table I to IV show the PSNR values for both the methods and normal correlation values.

Table IV: PSNR and NC for DWT-SVD with attacks

| PSNR VALUES | ATTACK | NC |
|---|---|---|
| 44.2012 | MEDIAN | 0.5622 |
| 45.1190 | NOISE | 0.4797 |
| 42.5848 | ROTATION | 0.5558 |
| 51.2546 | SHEAR | 0.5802 |
| 50.8055 | CROP | 0.6810 |
| 46.6372 | MEDIAN | 0.5862 |
| 43.0819 | NOISE | 0.4992 |
| 46.7991 | ROTATION | 0.5725 |
| 48.1555 | SHEAR | 0.6461 |
| 53.2543 | CROP | 0.7753 |

## VI. CONCLUSION

In this paper watermarking algorithm for medical images for securing such as authentication, integrity etc has proposed. This paper uses blind watermarking techniques such as DWT-SVD and DWT-DCT-SVD both, results are evaluated for respective techniques. It has been found that DWT-DCT-SVD based watermarking algorithm is robust when compared with DWT-SVD method. This method can be used for authentication and data hiding purposes. The future work can be extended to replace DWT method by discrete curvelet transform to improve the robustness of the watermark.